\documentclass[11pt]{article}
\usepackage{amssymb}
\usepackage[dvips]{graphicx,color}
\def\sint{\int \!\!\!\!\!\!\!\! \sum_{p}}

\def\tr{\mbox{tr}\,}
\def\Tr{\mbox{Tr}\,}

\def\openone{\leavevmode\hbox{\small1\kern-3.8pt\normalsize1}}%

\def\mf{{\mbox{\tiny\em MF}}}
\def\vol{{\mbox{\tiny\em V}}}
\def\qq{\langle\bar qq\rangle}

\def\be{\begin{equation}}
\def\ee{\end{equation}}
\def\bea{\begin{eqnarray}}
\def\eea{\end{eqnarray}}
\renewcommand{\thesection}{\Roman{section}}
\renewcommand{\thefootnote}{$\dagger$}

\begin{document}
\thispagestyle{empty}
\begin{center}
\vspace*{-1cm}
{\Large \bf Characteristics of the chiral phase transition \\
\rule{0cm}{.59cm} in nonlocal quark models}
\vspace*{.8cm} \\
{ D. G\'omez Dumm$^{a,b}$ and  N.N. Scoccola$^{b,c,d}$}
\vspace*{0.4cm} \\
$^a$ {\em IFLP, CONICET $-$ Depto.\ de F\'{\i}sica,
     Universidad Nacional de La Plata, \\
     C.C. 67, 1900 La Plata, Argentina.} \\
$^b$ {\em CONICET, Rivadavia 1917, 1033 Buenos Aires, Argentina.}\\
$^c$ {\em Physics Department, Comisi\'on Nacional de Energ\'{\i}a At\'omica, \\
 Av.Libertador 8250, 1429 Buenos Aires, Argentina.} \\
$^d$ {\em Universidad Favaloro, Sol{\'\i}s 453, 1078 Buenos Aires,
Argentina}
\vspace*{.5cm}
\date{\today}
\begin{abstract}
The characteristics of the chiral phase transition are analyzed within the
framework of chiral quark models with nonlocal interactions in the mean
field approximation. In the chiral limit, we develop a semi-analytic
framework which allows us to explicitly determine the phase transition
curve, the position of the critical points, some relevant critical
exponents, etc. For the case of finite current quark masses, we show the
behavior of various thermodynamical and chiral response functions across
the phase transition.
\end{abstract}
\vspace*{.3cm}
\end{center}

\vspace{1cm}

\renewcommand{\thefootnote}{\arabic{footnote}}
\setcounter{footnote}{0}

\section{Introduction}

The behavior of strongly interacting matter under extreme conditions of
temperature and/or density has important consequences in nuclear and
particle physics as well as in astrophysics and cosmology. From the
theoretical point of view, even if a significant progress has been made on
the development of ab initio calculations as lattice
QCD~\cite{All03,Fod04,Kar03}, these are not yet able to provide a detailed
knowledge of the full QCD phase diagram, and most theoretical approaches
rely in the study of low energy effective models. Qualitatively we expect
that chiral symmetry, which is broken at very low temperatures and
densities, will be restored as the temperature and/or density are
increased. However, the precise characteristics of this phase transition
are still not known. For two massless flavors most effective approaches to
QCD suggest the existence of a tricritical point on the $(T,\mu)$ plane,
which separates a first order phase transition line found at lower $T$ and
larger $\mu$, and a second order transition line where the chiral
restoration occurs for higher $T$ and lower $\mu$. For two light flavors a
similar behavior is expected, although the second order transition line is
replaced by a more or less sharp crossover and, correspondingly, the
tricritical point becomes an end point. In any case, for a given effective
model that can provide a reasonable successful description of low energy
strong interactions, it is very important to obtain as much information as
possible about the characteristics of the chiral restoration transition.
In previous works~\cite{GDS00,DS02} we have begun the study of chiral
restoration in the context of chiral quark models with nonlocal
interactions~\cite{Rip97}, which can be considered as some nonlocal
extension of the widely studied Nambu$-$Jona-Lasinio model~\cite{reports}.
In fact, nonlocality arises naturally in the context of several successful
approaches to low-energy quark dynamics as, for example, the instanton
liquid model~\cite{SS98} and the Schwinger-Dyson resummation
techniques~\cite{RW94}. It has been also argued that nonlocal covariant
extensions of the NJL model have several advantages over the local scheme
like e.g.\ a natural regularization scheme which automatically preserves
the anomalies~\cite{AS99}, small NLO corrections~\cite{Rip00}, etc. In
addition, it has been argued~\cite{Sti87,BB95} that a proper choice of the
nonlocal regulator and the model parameters can lead to some form of quark
confinement, in the sense that the effective quark propagator has no poles
at real energies.

Several studies~\cite{BB95,SDS04,BGR02} have shown that these nonlocal
chiral quark models provide a satisfactory description of the hadron
properties at zero temperature and density. The aim of the present work is
to complement the analysis of Refs.~\cite{GDS00,DS02}, presenting further
details about the chiral phase transition within these schemes. Indeed, we
show that in the chiral limit it is possible to develop a semi-analytic
framework which allows us to explicitly determine the phase transition
curve, the position of the critical points, some relevant critical
exponents, etc. For the case of finite current quark masses, we present
the behavior of various thermodynamical and chiral response functions
across the phase transition. In particular, it is found that thermal and
chiral susceptibilities show clear peaks which allow to define a phase
transition temperature as it is usually done in lattice calculations.

The paper is organized as follows. In Sect.\ II we provide a short
description of the model and its treatment in the mean field approximation
(MFA). In Sect.\ III we study the phase transition in the chiral limit by
performing the Landau expansion of the free energy. We also obtain the MFA
critical exponents. In Sect.\ IV we present and discuss the behavior of
the different thermodynamical and chiral response functions for the case
of finite current quark masses, and the corresponding phase diagrams are
shown. In Sect.\ V we present a summary of our main results and
conclusions. Finally, some details of the calculations are given in
Appendices A and B.

\section{Nonlocal chiral quark models}

Let us begin by stating the Euclidean action for the nonlocal chiral quark model in
the case of two light flavors and $SU(2)$ isospin symmetry. One has
\footnote{For simplicity we neglect here possible diquark channels.
See Ref.\cite{DGS04} for details on their r\^ole in the phase diagram
of this type of models.}
\begin{equation}
S_E = \int d^4 x \left[ \bar \psi (x) \left(- i \rlap/\partial  + m_c\,
\openone \right) \psi (x) - \frac{G}{2}\, j_a(x) j_a(x) \right] \ ,
\label{action}
\end{equation}
where $\psi = (u\ d)^T$ and $m_c$ stands for the $u$ and $d$ current quark
mass. The current $j_a(x)$ is given by
\begin{equation}
j_a (x) = \int d^4 y\ d^4 z \ \tilde r(y-x) \ \tilde r(x-z) \
\bar \psi(y)\,\Gamma_a \,\psi(z)\,,
\end{equation}
where $\Gamma_a = ( \openone, i \gamma_5 \vec \tau )$, and the function
$\tilde r(x-y)$ is a nonlocal regulator. The latter can be translated into
momentum space,
\begin{equation}
\tilde r(x-z) = \int \frac{d^4p}{(2\pi)^4} \ e^{-i(x-z) p} \ r(p) \;.
\end{equation}
In fact, dimensional analysis together with Lorentz invariance implies that $r(p)$
can only be a function of $p^2/\Lambda^2$, where $\Lambda$ is a cutoff parameter
describing the range of the nonlocality in momentum space. Hence we will use for
the Fourier transform of the regulator the form $r_\Lambda(p^2)$ from now on.

{}From the Euclidean action in Eq.~(\ref{action}), the partition function
for the model at zero $T$ and $\mu$ is defined as
\begin{equation}
{\cal Z}_0 = \int {\cal D} \bar\psi {\cal D} \psi \  e^{-S_E}\;.
\label{zcero}
\end{equation}
We perform now a standard bosonization of the theory, introducing the
sigma and pion meson fields $M_a(x) = (\sigma(x), \vec \pi(x))$. In this
way the partition function can be written as~\cite{DS02}
\begin{equation}
{\cal Z}_0 = \int {\cal D} \sigma {\cal D} \pi \  \det A(M_a)\
\exp \left[-\frac{1}{2 G} \int \frac{d^4 p}{(2\pi)^4} \ M_a^2(p) \right]\,,
\label{z}
\end{equation}
where the operator $A$ reads in momentum space
\begin{equation}
A(M_a) = (\,-\rlap/p + m_c)\,(2\pi)^4 \,\delta^{(4)}(p-p')
+ r_\Lambda(p^2)\, M_a(p-p')\,r_\Lambda({p'}^2)\, \Gamma_a\;.
\end{equation}

In what follows we will work within the mean field approximation, in which
the meson fields are expanded around their translational invariant vacuum
expectation values
\begin{eqnarray}
\sigma(x) &=& \bar \sigma +\  \delta \sigma (x) \\
\pi_i(x) &=& \delta \pi_i (x)
\end{eqnarray}
and the fluctuations $\delta \sigma (x)$ and $\delta \pi_i (x)$ are
neglected (vacuum expectation values of the pion fields vanish owing to
parity conservation). Within this approximation the determinant in
(\ref{z}) is formally given by
\begin{equation}
\det A = \exp \Tr \log A = \exp V^{(4)} \int \frac{d^4 p}{(2\pi)^4}\;
\tr \log \left[\,-\rlap/p + m_c + \bar \sigma \, r_\Lambda^2(p^2)\,\right]
 \,,
\label{det}
\end{equation}
where $\tr$ stands for the trace over the Dirac, flavor and color indices,
and $V^{(4)}$ is the four-dimensional volume of the path integral.

\section{Phase transition in the chiral limit}

In Refs.~\cite{GDS00,DS02} we have analyzed the chiral restoration within
nonlocal chiral quark models for some definite regulators. In particular,
we have shown that in the chiral limit this phase transition is a second
order one for low values of the chemical potential $\mu$. In this Section
we will rederive this result following the so-called classical approach
proposed by Landau, in which one expands the free energy in powers of the
order parameter (in this case the quark condensate $\langle\bar
qq\rangle$) in the vicinity of the critical temperature. This shows
the equivalence of the chiral restoration in nonlocal chiral quark models
with the corresponding phase transitions taking place in other physical
systems, such as ferromagnets or superfluids. As stated, we will work
within the mean field approximation, which in this context means to
approximate the path integral in Eq.\ (\ref{z}) by its maximum, reached at
some saddle point.

In our case, the partition function in the grand canonical ensemble for
finite temperature $T$ and chemical potential $\mu$ can be obtained from
Eqs.~(\ref{z}) and (\ref{det}). In these expressions, the integrals over
four-momentum space have to be replaced by Matsubara sums according to
\begin{equation}
\int \frac{d^4 p}{(2\pi)^4}\; F(p_4,\vec p) \quad \to \quad
\sint F(p_4,\vec p) \equiv
T \sum_{n=-\infty}^\infty \int \frac{d^3 p}{(2\pi)^3}\;
F(\omega_n - i\mu,\vec p)\;,
\end{equation}
where $\omega_n$ are the Matsubara frequencies corresponding to fermionic
modes, $\omega_n = (2 n+1) \pi T$. In the same way the volume $V^{(4)}$ is
replaced by $V/T$, $V$ being the three-dimensional volume in coordinate
space. As in Refs.~\cite{GDS00,DS02}, we are assuming here that the quark
interactions only depend on the temperature and chemical potential through
the arguments of the regulators. The grand canonical thermodynamical
potential per unit volume is thus given by \cite{DS02}
\begin{eqnarray}
\omega_\mf (T,\mu,m_c) & = & - \frac{T}{V} \, \log {\cal
Z}_\mf (T,\mu,m_c)
\nonumber \\
& = & \frac{\bar \sigma^2}{2 G}\;
- 4 N_c \sint \log \left[ p^2 + \Sigma^2(p^2) \right] \ ,
\label{omega}
\end{eqnarray}
where $\Sigma(p^2) = m_c + \bar \sigma \, r^2(p^2)$ stands for the quark
selfenergy, and the mean field value $\bar\sigma(T,\mu,m_c)$ is obtained
from the condition
\begin{equation}
\frac{\partial \omega_\mf}{\partial \bar \sigma} = 0 \ .
\label{gap}
\end{equation}
In fact, $\omega_\mf$ turns out to be divergent. The regularization
procedure used here amounts to define
\begin{equation}
\omega_\mf^{(reg)}(T,\mu,m_c) =
\omega_\mf(T,\mu,m_c)-\omega_{free}(T,\mu,m_c)
+\omega_{free}^{(reg)}(T,\mu,m_c)+\omega_0\ ,
\label{omegareg}
\end{equation}
where $\omega_{free}^{(reg)}(T,\mu,m_c)$ is the regularized expression for
the thermodynamical potential of a free fermion gas, and $\omega_0$ is a
constant fixed by the condition $\omega_\mf^{(reg)}=0$ at $T=\mu=0$
(see Appendix A for details).

In the analogy with a ferromagnetic system, the chiral condensate can be
identified with the magnetization per unit volume, $\qq\longleftrightarrow
-M/V$, whereas the current quark mass $m_c$ plays the r\^ole of the
external magnetic field $H$. One has
\begin{equation}
\qq = \frac{1}{2} \left(\frac{\partial\omega}{\partial
m_c}\right)_{T,\mu}\;,
\label{qq}
\end{equation}
where the $\frac12$ factor results from the fact that this relation holds
for each quark flavor separately, i.e.\ $q=u,d$, while they have a common
mass $m_c$. In the chiral limit, the existence of a second order phase
transition for a fixed value of $\mu$ implies that the condensate $\qq$
goes to zero when the temperature $T$ approaches from below a given
critical value $T_c(\mu)$, above which one has $\qq=0$ and the chiral
symmetry is restored. Thus, for $T\sim T_c(\mu)$ and to leading order in
$m_c$,  one can perform the Landau expansion (see Appendix A)
\begin{eqnarray}
\omega^{(reg)}_\mf(T,\mu,m_c) & = & \omega_0 \ + \
\omega^{(reg)}_{free}(T,\mu,m_c=0) \
+  \ A(T,\mu)\, \qq^2 + \ C(T,\mu)\, \qq^4 \ + \nonumber \\
& & \qquad \qquad  + \ 2\, m_c\, \qq \  + \ {\cal O}(\qq^6,\qq^3\,m_c,\,m_c^2) \ ,
\label{expansion}
\end{eqnarray}
where the coefficients $A$ and $C$ are given by
\begin{eqnarray}
A(T,\mu) & = & \frac{1}{4\,N_c^2\, S_{11}^2(T,\mu)}\,
\left[\frac{1}{8G} - N_c\, S_{21}(T,\mu)\right] \nonumber \\
C(T,\mu) & = & \frac{S_{42}(T,\mu)}
{128\,N_c^3\, S_{11}^4(T,\mu)}\ - \
\frac{S_{32}(T,\mu)}{32\,N_c^4\, S_{11}^5(T,\mu)}\,
\left[\frac{1}{8G} - N_c\, S_{21}(T,\mu)\right]\;,
\end{eqnarray}
with
\begin{equation}
S_{mn}(T,\mu)\ =\ \sint \frac{r_\Lambda^{2m}(p^2)}{p^{2n}}\ \ .
\end{equation}
The regularized thermodynamical potential for a massless fermion gas
---second term on the r.h.s.\ of Eq.~(\ref{expansion})--- can be calculated
by evaluating the integral in Eq.~(\ref{ofreeint}) in the massless case.
One obtains
\begin{equation}
\omega_{free}^{(reg)}(T,\mu,0) = -\,\frac{N_c}{3}\left[\frac{7\pi^2}{30}\;T^4\,
+\,T^2\mu^2\,+\, \frac{1}{2\pi^2}\; \mu^4 \right]\,.
\label{omegafree}
\end{equation}

In the limit $m_c=0$, it can be seen~\cite{khuang} that for $C > 0$
the system undergoes a second order phase transition at a critical
temperature $T_c(\mu)$ obeying $A(T_c(\mu),\mu)=0$. This implies
\begin{equation}
S_{21}(T_c(\mu),\mu)\ =\ \frac{1}{8\,G\,N_c}\ ,
\label{tin}
\end{equation}
which defines a second order transition curve in the $(T,\mu)$ plane. As
it is described in Appendix B, for sufficiently small (but relevant)
values of $T$ and $\mu$ the Matsubara sum implicit in
$S_{21}(T_c(\mu),\mu)$ can be analytically worked out. This leads to a
simple relation between $T_c$ and $\mu$, namely
\begin{equation}
\frac{\pi^2}{3}\left(\frac{T_c(\mu)}{\Lambda}\right)^2 +
\left(\frac{\mu}{\Lambda}\right)^2 = \beta_0 - \frac{\pi^2}{N_c}
\frac{1}{G \Lambda^2} \ ,
\label{second}
\end{equation}
where $\beta_0 = \Lambda^{-2} \int\ dp\  p\ r_\Lambda^{4}(p^2)$ is a
dimensionless quantity which depends only on the shape of the regulator.
Note that this relation generalizes that obtained in Ref.~\cite{SKP99} for
the NJL model. We also point out that for a given value of $\mu$, and
temperatures which are close to the critical value $T_c(\mu)$ obtained
from Eq.~(\ref{tin}), one has
\begin{equation}
A(T,\mu) = \lambda\, t \ ,
\end{equation}
where
\begin{equation}
t\equiv (T-T_c(\mu))/T_c(\mu)\ \ , \hspace{1cm}
\lambda = \frac{T^2_c(\mu)}{48\,N_c S_{11}^2(T_c(\mu),\mu)}\ \ .
\end{equation}

We note in passing that once the Landau expansion has been established, it
is a usual textbook exercise~\cite{khuang} to derive the critical
exponents ruling the behavior of the specific heat, the order parameter
$\qq$ and the chiral susceptibility near the second order critical points,
\begin{equation}
c_{\vol,\mu,m_c}|_{m_c=0} \sim |t|^{-\alpha}\;,\quad \qq|_{m_c=0} \sim
|t|^\beta\;,\quad \chi_{\vol,T,\mu}|_{m_c=0} \sim |t|^{-\gamma}\;,\quad
\qq|_{t=0} \sim m_c^{\;\delta} \ .
\end{equation}
Here the specific heat and the chiral susceptibility are defined as
\begin{eqnarray}
c_{\vol,\mu,m_c} & = & -\; T
\left(\frac{\partial^2 \omega(T,\mu,m_c)}{\partial T^2} \right)_{\mu,m_c}
= \ T \left(\frac{\partial s(T,\mu,m_c)}{\partial T}\right)_{\mu,m_c}\ ,
\label{cvmu} \\
\chi_{\vol,T,\mu} & = & - \;\frac12\, \left(\frac{\partial^2
\omega(T,\mu,m_c)}{\partial m_c^2} \right)_{T,\mu} = \ -\,
\left(\frac{\partial\qq (T,\mu,m_c)}{\partial m_c} \right)_{T,\mu}\ ,
\label{chimu}
\end{eqnarray}
where $s(T,\mu,m_c)$ is the entropy density. As expected, one obtains the
mean field critical exponents\footnote{At the tricritical point, the mean
field critical exponents are $\alpha=\frac{1}{2}$, $\beta=\frac{1}{4}$,
$\gamma=1$ and $\delta=5$.}
\begin{equation}
\alpha = 0\;,\qquad\beta = {\textstyle \frac{1}{2}}
\;,\qquad\gamma = 1\;,\qquad\delta = 3\;,
\label{veinte}
\end{equation}
a result that was under discussion in the somewhat related Dyson-Schwinger
models of QCD~\cite{expo}. Eq.~(\ref{veinte}) implies that the chiral
susceptibility diverges at $T=T_c$. In the case of the specific heat,
there is no divergence but a finite discontinuity at $m_c=0$, $t=0$,
\begin{equation}
c_{\vol,\mu,m_c}\big|_{t=0^-} = c_{\vol,\mu,m_c}\big|_{t=0^+} +
\frac{N_c\, T_c^3}{36\,S_{42}}\; ,
\end{equation}
where $c_{\vol,\mu,m_c}|_{t=0^+}$ is the specific heat at constant $\mu$
due to the free fermion gas pressure
$p_{free}=-\omega_{free}^{(reg)}(T,\mu,0)$, evaluated at $T=T_c$. From
Eq.~(\ref{omegafree}) one gets
\begin{equation}
c_{\vol,\mu,m_c}\big|_{t>0} = \frac{2}{3}\; N_c \, T
\left(\mu^2+\frac{7\,\pi^2}{5}\ T^2\right)\, .
\label{cvfree}
\end{equation}

As stated, the transition remains second order as long as $C(T_c(\mu),\mu)
> 0$. This is expected to be the case for low values of the chemical
potential. However, if $\mu$ is increased, one could reach a point at
which the coefficient $C$ vanishes. Beyond this point, called
``tricritical'', the system undergoes a first order phase transition (the
order parameter $\qq$ is discontinuous at the corresponding critical
temperature). In this way, the tricritical point is obtained from the
conditions $A(T,\mu)=C(T,\mu)=0$, or equivalently
\begin{equation}
S_{42}(T_c(\mu),\mu) = 0\ .
\label{tricrit}
\end{equation}
Again,  for sufficiently small (but relevant) values of $T$ and $\mu$ the
Matsubara sum in $S_{42}$ can be analytically worked out (see Appendix B),
leading to
\begin{equation}
S_{42}(T,\mu) = \frac{1}{8\pi^2}\left[\;\beta_1\, (T/\Lambda)^2
\left(\frac{\pi^2}{3}\, + (\mu/T)^2\right)
+ \beta_2 - f(\mu/T) - \log (T/\Lambda)\right]\;,
\label{s42}
\end{equation}
where
\begin{eqnarray}
\beta_1 & = & -\, 8\;\Lambda^2\; \left.\frac{dr_\Lambda(p^2)}{dp^2}\right|_{p=0}
\ ,
\nonumber \\
\beta_2 & = & \int_0^\infty \frac{dp}{p}
\left[r_\Lambda^8(p^2)-e^{-4p^2}\right] + \frac{1}{2}(\gamma-1) - \log 2\pi\ ,
\end{eqnarray}
and $f(x)$, which satisfies $f(0)=0$, is given by Eq.~(\ref{fx}) of
Appendix B. It is interesting to notice the similarity between our
expressions in Eqs.~(\ref{second}) and (\ref{tricrit}-\ref{s42}) and
those obtained in Ref.~\cite{zsolt} within a very different theoretical
approach. A brief discussion on the subject is also included in Appendix
B.

\hfill

Let us consider for definiteness a nonlocal model in which the regulator
is a Gaussian function,
\begin{equation}
r_\Lambda(p^2)=\exp (-p^2/2\Lambda^2)\ .
\end{equation}
In this case one has $\beta_1=4$, while the integral in $\beta_2$
vanishes. In addition, it is easily seen that $S_{42}(T,0)$ is a positive
function of $T/\Lambda$, which implies that in the chiral limit the model
leads to a second order chiral phase transition for vanishing chemical
potential. The transition remains second order when $\mu$ is increased up
to the tricritical point, and the corresponding transition line in the
$(T,\mu)$ plane can be immediately obtained from Eq.~(\ref{second}), with
$\beta_0=1/4$. The critical temperature at $\mu=0$ is thus given by
\begin{equation}
T_c(0)\ = \ \frac{\sqrt{3}}{2\pi}\ \Lambda \ \bigg( 1- \frac{4 \pi^2}{N_c
\, G\Lambda^2} \bigg)^{1/2} \ .
\label{tinto}
\end{equation}
We illustrate these features by considering a particular parameter set,
namely $\Lambda = 760$ MeV and $G=30$ GeV$^{-2}$. These are the parameters
corresponding to Set II in Ref.\ \cite{DS02}, leading to $T_c(0)= 102$
MeV. The second order phase transition is clearly shown in Fig.\ 1(a),
where the solid line represents the chiral condensate for $\mu=0$ as a
function of $T$. Now using Eq.~(\ref{tricrit}) one can find the position
of the tricritical point, which in this case is found to be located at
$(T,\mu)=$(72 MeV,133 MeV). In Fig.\ 1(a), dashed and dashed-dotted lines
show the behavior of the chiral condensate for $\mu=100$ MeV and $\mu=200$
MeV, below and above the tricritical point respectively. Both the values
for $T_c(0)$ and the position of the tricritical point are found to be in
very good agreement with those numerically obtained in Ref.~\cite{DS02}.
Finally, the described effect on the specific heat is shown in Fig.\ 1(b),
where we plot $c_{\vol,\mu}$ as a function of the temperature for $\mu=0$,
100 MeV and 200 MeV.

\begin{figure}[htb]
\centerline{
   \includegraphics[height=6.5truecm]{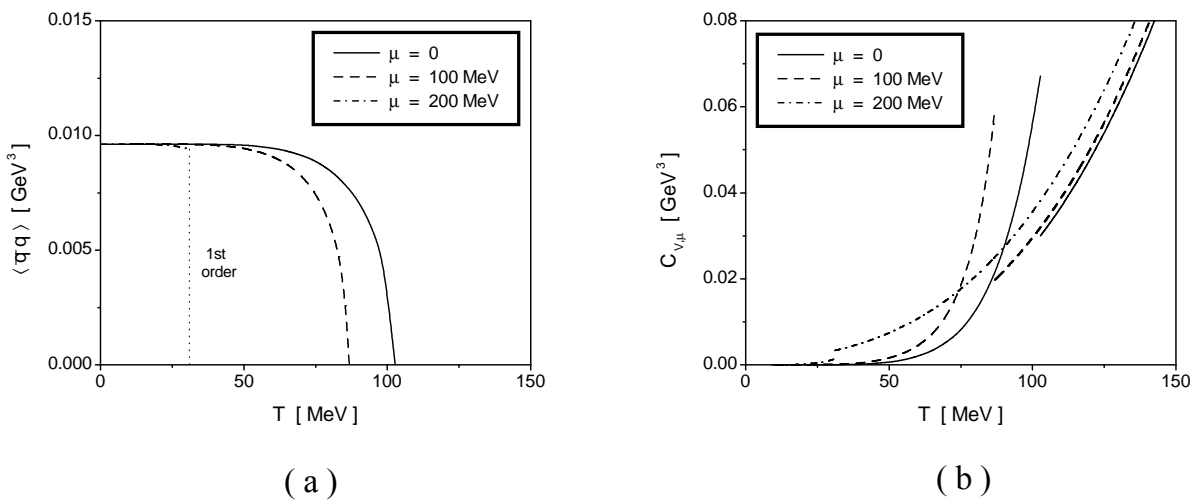}
   } \caption{(a) Chiral condensate vs.\ temperature, and (b) specific
heat $c_{\vol,\mu}$ vs.\ temperature. Solid, dashed and dashed-dotted
lines correspond to $\mu=0$, $\mu=100$ MeV and $\mu=200$ MeV
respectively.}
\end{figure}

The result in Eq.~(\ref{tinto}) is useful to point out a generic feature
of this type of models in the chiral limit. In fact, it is not difficult
to show~\cite{BB95} that, in such limit, for zero $T$ and $\mu$ one
obtains expressions for the pion decay constant $f_\pi^{(0)}$ and the
chiral condensate $\qq^{(0)}$ such that $f_\pi^{(0)}/\Lambda$ and
$\qq^{(0)}/\Lambda$ depend only on the combination $G \Lambda^2$. In order
to get a ratio $f_\pi^{(0)}/\qq^{(0)} \simeq 0.4$, as required by
phenomenology, these expressions lead to $G\Lambda^2\simeq 16$. If this
value is now replaced in Eq.~(\ref{tinto}) one gets $T_c(0)/\Lambda \simeq
0.12$, which is somewhat large, but still lies within the expected range
of validity. However, if one imposes the critical temperature $T_c(0)$ to
be approximately equal to 170 MeV, as suggested by lattice QCD
calculations, one gets $\Lambda \simeq 1.4$ GeV, which, together with $G
\Lambda^2 \simeq 16$, enhances $f_\pi^{(0)}$ and $\qq^{(0)}$ up to roughly
40\% above the corresponding phenomenological values. Thus, it can be
generically said that, if in the chiral limit ---and within the mean field
approximation--- one wants to satisfy the phenomenological constraints on
the values of $f_\pi$ and $\qq$, the nonlocal models under consideration
lead to a relatively low critical temperature $T_c(0)$. Of course, this
might be modified by finite quark mass effects and/or beyond MFA
corrections.

\section{Phase transition and response functions for finite quark mass}

In this Section we concentrate in the analysis of the phase transition for
finite current quark masses in the isospin limit. For this purpose we will
consider different thermodynamical response functions. In particular, it
will be seen that, for reasonable values of the current quark mass $m_c$,
the chiral susceptibilities show clear peaks that can be used to define
the transition curves in the crossover region.

The response functions are basically given by the second derivatives of
the Helmholtz free energy $F(T,V,\bar N)$. The latter is related with the
thermodynamical potential by
\begin{equation}
F(V,T,\bar N) = \left[ \Omega(V,T,\mu) + 2\,\bar N \;\mu
\right]_{\mu=\mu(V,T,\bar N)}\ ,
\end{equation}
where $\bar N = - \frac12\;( \partial \Omega/ \partial \mu)_{V,T}$ (as in
the case of the quark condensate, the $\frac12$ factor comes from defining
$\bar N = \bar N_u=\bar N_d$). We will consider as before the limit of a
large system in which $\Omega(V,T,\mu) = V\,\omega(T,\mu)$, $F(V,T,\bar N)
= V\,f(T,\bar N/V)$, thus instead of average particle number $\bar N$ our
quantities will be given in terms of the average particle density $\rho =
\bar N/V$. One can define three independent thermodynamical response
functions, namely the specific heat at fixed volume and particle number
$c_{\vol,\rho,m}$, the isothermal compressibility $\kappa_{T,\rho,m_c}$
and the coefficient of thermal expansion $\alpha_{p,\rho,m_c}$. These are
given by
\begin{eqnarray}
c_{\vol,\rho,m_c} & = & - \;\frac{T}{V}
\left(\frac{\partial^2 F(T,V,\bar N,m_c)}{\partial T^2} \right)_{V,\bar N,m_c}
= T \left(\frac{\partial s(T,\rho,m_c)}{\partial T} \right)_{\rho,m_c}
\label{cvrho} \\
\frac{1}{\kappa_{T,\rho,m_c}} & = & V \left(\frac{\partial^2 F(T,V,\bar
N,m_c)}{\partial V^2} \right)_{T,\bar N,m_c} = \rho \left(\frac{\partial
p(T,\rho,m_c)}{\partial \rho} \right)_{T,m_c} \\
\alpha_{p,\rho,m_c} & = &
\frac{1}{V}\left(\frac{\partial V}{\partial T}\right)_{p,\bar N,m_c} =
-\; \kappa_{T,\rho,m_c} \left(\frac{\partial^2
F(T,V,\bar N,m_c)}{\partial V\;\partial T} \right)_{\bar N,m_c}
\nonumber \\
&  = & \kappa_{T,\rho,m_c} \left(\frac{\partial p(T,\rho,m_c)}{\partial T}
\right)_{\rho,m_c}
\label{respfun}
\end{eqnarray}
There are also other thermodynamical response functions that can be
defined, such as the specific heat at constant pressure $c_{p,\rho,m_c}$
and the adiabatic compressibility $\kappa_{s,\rho,m_c}$. However, these
can be written in terms of the three quantities in
Eqs.~(\ref{cvrho}-\ref{respfun}).

Notice that we have explicitly included the dependence of the free energy
on the current quark mass $m_c$. In fact, one can still define some extra
response functions associated with the chiral transition by differentiating the
free energy with respect to $m_c$. In the analogy with magnetic systems,
these would be the specific heats at constant magnetization and constant
applied magnetic field, $C_M$, $C_H$,
the isothermal and adiabatic susceptibilities $\chi_T$, $\chi_S$,
and the coefficient of thermal magnetization, $\alpha_H$.
As before, these quantities are related to each other, leaving only two
new independent response functions. Here we choose to consider those
analogous to the isothermal susceptibility and the coefficient of thermal
magnetization, thus we define
\begin{eqnarray}
\chi_{\vol,T,\rho} & = & - \;\frac{1}{2\,V}\;
\left(\frac{\partial^2 F(T,V,\bar N,m_c)}{\partial m_c^2} \right)_{T,V,\bar N}
= -\; \left(\frac{\partial\qq (T,\rho,m_c)}{\partial m_c} \right)_{T,\rho}
\label{chirho} \\
\alpha_{\vol,\rho,m_c} & = & -\;\frac{1}{2\,V}\;\left(\frac{\partial^2
F(T,V,\bar N,m_c)}{\partial m_c\;\partial T} \right)_{V,\bar N} =
-\; \left(\frac{\partial\qq (T,\rho,m_c)}{\partial T}\right)_{\rho,m_c}\ .
\label{alpharho}
\end{eqnarray}

Let us show the numerical results for these quantities for the case of a
nonlocal Gaussian regulator. We have chosen the parameter set
$\Lambda=760$~MeV, $G=30$~GeV$^{-2}$ and $m_c=7.7$~MeV, which corresponds
to Set II in the notation of Ref.~\cite{DS02}. It can be seen that these
parameters lead to the empirical values of the pion mass and decay
constant, and provide reasonable results for the chiral quark condensate
and the quark selfenergy $\Sigma(0)$ at zero $T$ and $\mu$. In Fig.~2 we
show the curves corresponding to the specific heat $c_{\vol,\rho,m_c}$ and
the chiral response functions $\chi_{\vol,T,\rho}$ and
$\alpha_{\vol,\rho,m_c}$. In the left panels, these functions are plotted
versus the temperature for three representative values of the density (the
quoted values are referred to nuclear matter density, $\rho_0\simeq 1.3
\times 10^6$ MeV$^3$). Notice that for low nonzero densities the system is
homogeneous only for temperatures which exceed a critical value. Below
this limit, as we will see, there is a region where phases with broken and
restored chiral symmetry coexist. In the right panels, the same response
functions are plotted as functions of the density, fixing the temperature
at 50~MeV and 100~MeV. Once again, for low temperatures there is a mixed
phase region, and as a consequence the functions are not well defined at
intermediate densities. We have also analyzed the response functions
corresponding to the parameter set given by $G=50$~GeV$^{-2}$,
$m_c=10.5$~MeV and $\Lambda=627$~MeV, or Set I in the notation of
Ref.~\cite{DS02}. Set I and II might be interpreted as confining and
nonconfining respectively, where confinement is understood in the sense
that the pole structure of the quark propagator does not allow quarks to
materialize on-shell in Minkowski space~\cite{Sti87,BB95}. The curves in
the case of Set I do not differ qualitatively from those shown in Fig.~2,
therefore they have not been included here.

\begin{figure}[htb]
\centerline{
   \includegraphics[height=15truecm]{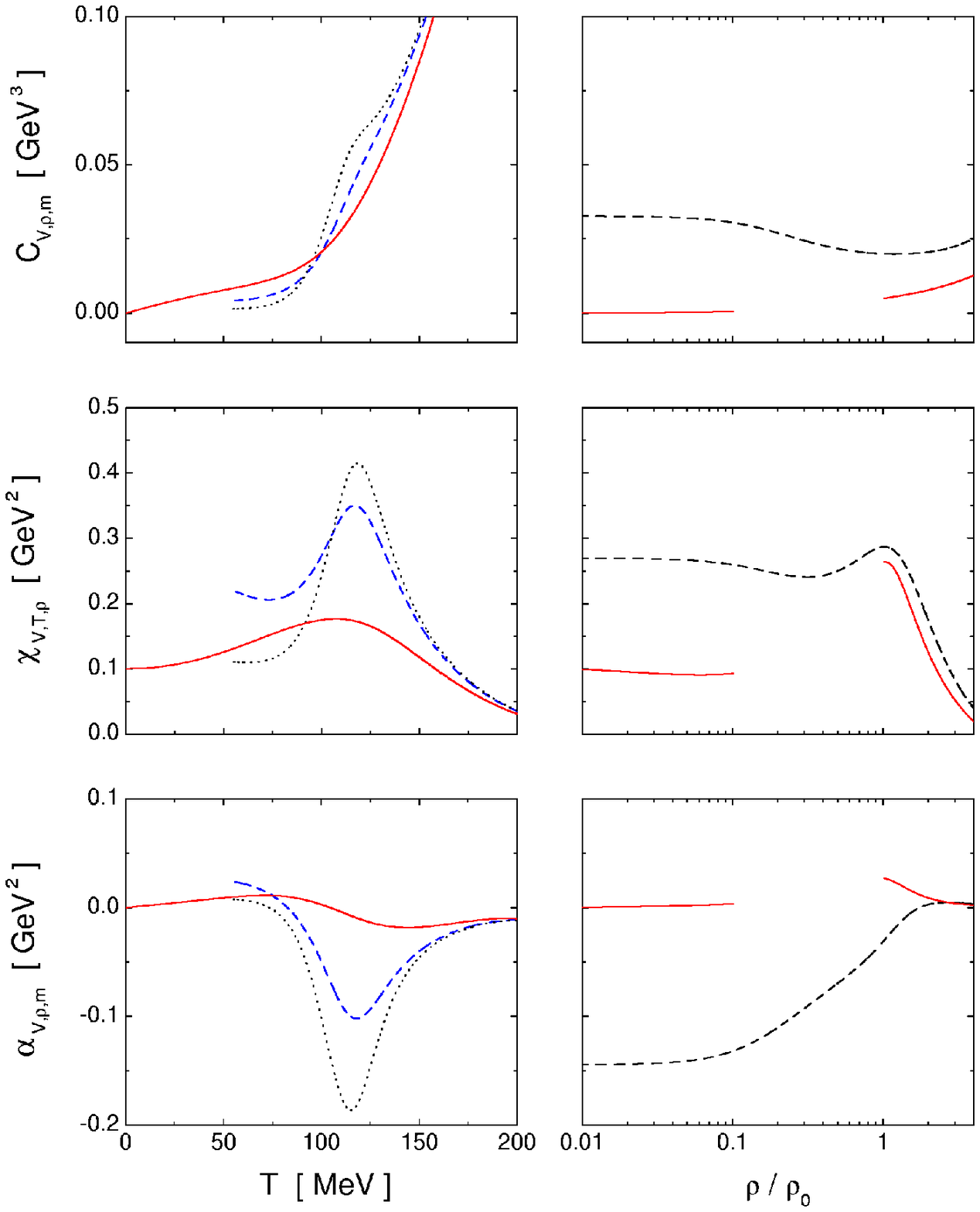}
   } \caption{Some response functions for fixed $\rho$, corresponding to
parameter Set II. In the left panels curves are given as functions of $T$,
with $\rho/\rho_0$ fixed at three representative values, namely 0.25
(dotted), 0.75 (dashed) and 2.0 (solid). Right panels show the curves as
functions of $\rho/\rho_0$ for $T$ fixed at 50 MeV (solid) and 100 MeV
(dashed).}
\end{figure}

Now let us pay special attention to the responses of the chiral condensate
(order parameter of the phase transition) at fixed temperature and chiral
quark mass (or ``magnetization''). These are given by Eqs.~(\ref{chirho})
and (\ref{alpharho}) and their behavior as functions of the temperature
is shown in the second and third rows of Fig.~2 (left panels). As it is
well known, susceptibilities are particularly useful to analyze the phase
transition features. In the present model, as shown in previous
works~\cite{GDS00,DS02}, for low temperatures the system undergoes a first
order chiral phase transition, which turns into a smooth crossover for
temperatures exceeding a given ``end point''. In this crossover region,
the transition is characterized by the presence of respective peaks in the
mentioned susceptibilities, the height and sharpness of these peaks giving
a measure of the crossover steepness. The same can be observed if one
looks at the chiral and thermal susceptibilities for fixed $\mu$, which
are given by the second derivatives of the thermodynamical potential, and
are in fact the natural quantities to deal with in the grand canonical
ensemble. The definition of the chiral susceptibility $\chi_{\vol,T,\mu}$
has been already given in Eq.~(\ref{chimu}), while the thermal
susceptibility at constant $\mu$ is defined as
\begin{equation}
\alpha_{\vol,\mu,m_c} \ = \ -\;\frac12\;
\left(\frac{\partial^2 \omega(T,\mu,m_c)}{\partial m_c\;\partial T} \right)_\mu
= -\,\left(\frac{\partial\qq (T,\mu,m_c)}{\partial T}\right)_{\mu,m_c}\ .
\end{equation}

The curves showing the behavior of the susceptibilities
$\chi_{\vol,T,\mu}$ and $\alpha_{\vol,\mu,m}$ as functions of the
temperature and the chemical potential are shown in Fig.~3 (the chosen
parameter set is the same as in Fig.~2). For completeness we also include
the curves for the specific heat at constant $\mu$, previously introduced
in Sect.~III (where the case of varying $m_c$ was discussed).
\begin{figure}[htb]
\centerline{
   \includegraphics[height=15truecm]{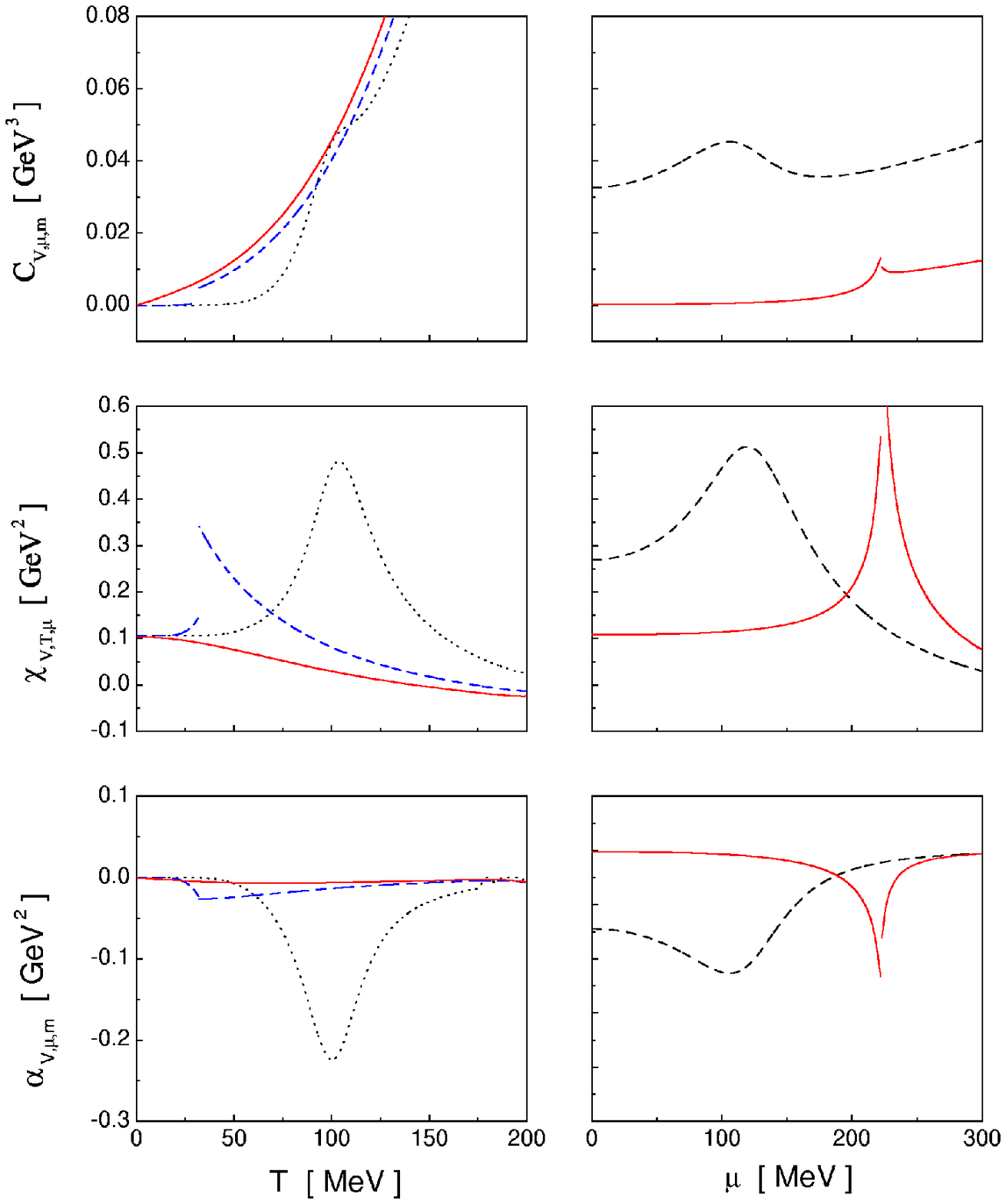}
   } \caption{Some response functions for fixed $\mu$, corresponding to
parameter Set II. In the left panels curves are given as functions of $T$,
with $\mu$ fixed at three representative values, namely 100 MeV (dotted),
250 MeV (dashed) and 300 MeV (solid). Right panels show the curves as
functions of $\mu$ for $T$ fixed at 50 MeV (solid) and 100 MeV (dashed).}
\end{figure}
In a natural way, the peaks in the curves for $\chi_{\vol,T,\mu}$ and
$\alpha_{\vol,\mu,m_c}$ can be used to define the position at which the
chiral transition occurs. Thus, one can extend the phase space diagram to
include the crossover transition curves in addition to the first order
ones. This is represented in Fig.~4, where crossover curves obtained from
the peaks in $\chi_{\vol,T,\mu}$ and $\alpha_{\vol,\mu,m_c}$ have been
represented by dashed and dotted lines respectively. Although chiral and
thermal susceptibilities lead to slightly different points, it is seen
that the transition region is well defined. Full lines correspond to the
first order phase transition, while the fat dots indicate the end points.
Left and right panels show the results for parameter Sets I and II
respectively, while upper (lower) panels show the phase space transition
curves in the $T-\mu$ ($T-\rho$) plane. Notice that in the $T-\rho$ phase
diagrams there is a region below the first order transition lines where
both phases are allowed. The latter can be interpreted~\cite{BR98} as a
zone in which droplets containing light quarks of mass $m_c$ coexist with
a gas of constituent, massive quarks. The dotted lines inside this zone
are the spinodals, i.e.\ the boundaries of the region in which the
energetically unfavored solutions can exist as metastable states.

Our results for the phase transition curves in the $T-\mu$ plane are
qualitatively similar to those obtained from lattice QCD
calculations~\cite{Fod04,Kar03}, although both the critical temperature at
$\mu=0$ and the end point temperature turn out to be relatively low in our
case (see discussion at the end of Sect.~III). For the sake of comparison,
it is also interesting to analyze the curvature of the phase boundary at
$\mu=0$. This is an appropriate quantity to be studied in lattice QCD,
where the main problem in the analysis of the $T-\mu$ phase diagram is the
inclusion of a finite real chemical potential. Recent lattice
calculations~\cite{All03} yield $T_c (d^2T_c/d\mu^2)|_{\mu=0}=-0.14\pm
0.06$, while the result obtained within the standard NJL model (up to some
finite current quark mass corrections) gives a value of  about $-0.40$
~\cite{Bub04}. In our model, the curvature can be calculated numerically
from the results plotted in Fig.~4, leading to values of about $-0.26$ and
$-0.30$ for Sets I and II, respectively. In the chiral limit the
corresponding calculation can be carried out from the analytical
expression in Eq.~(\ref{second}), leading to $T_c
(d^2T_c/d\mu^2)|_{\mu=0}=-3/\pi^2\simeq -0.304$.

\begin{figure}[htb]
\centerline{
   \includegraphics[height=12truecm]{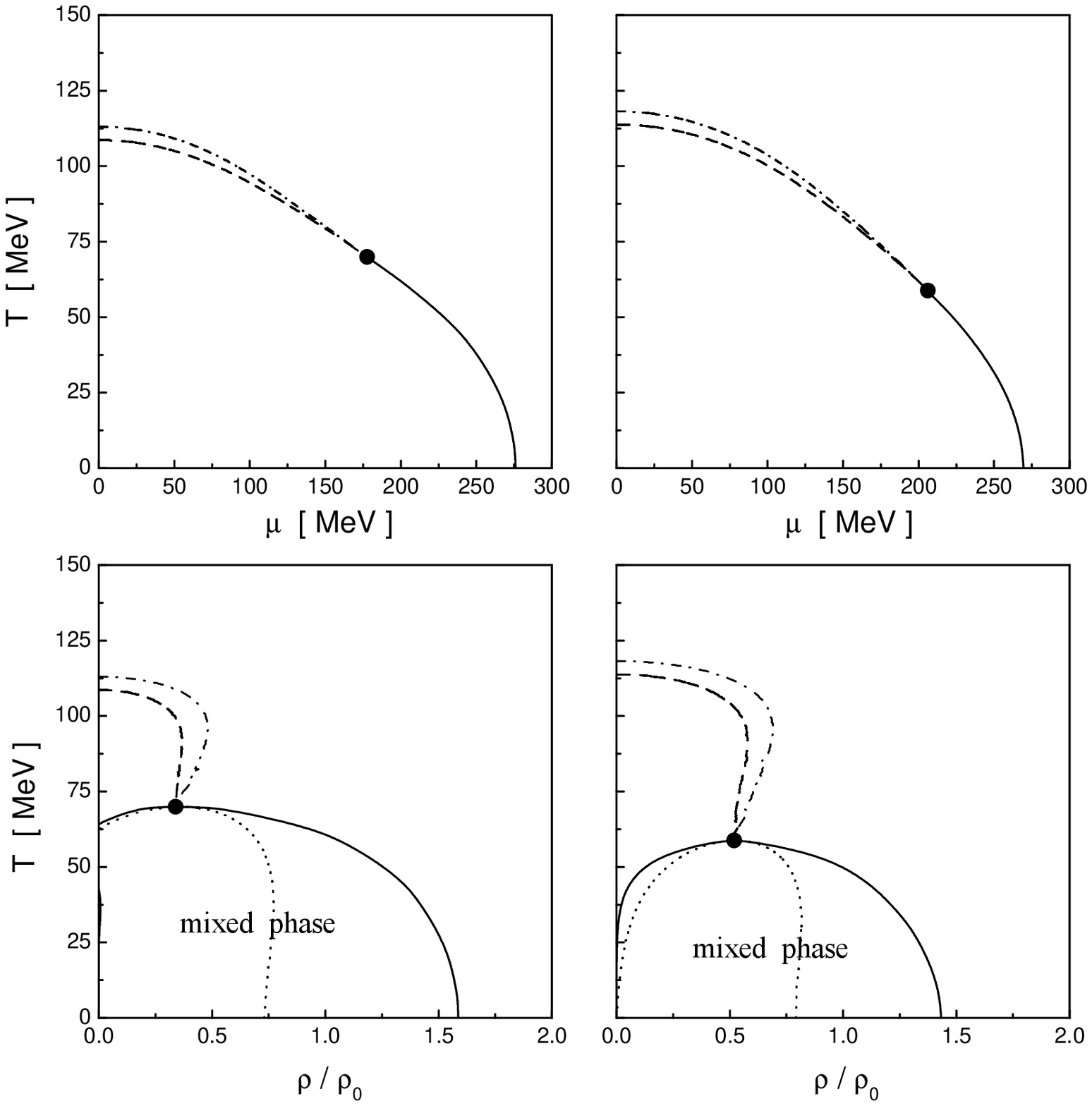}
   } \caption{Phase diagram showing the chiral transition curves in the
$T-\mu$ plane (upper panels) and the $T-\rho$ plane (lower panels), for
parameter Sets I (left) and II (right). Solid lines correspond to first
order phase transition curves, while dashed (dashed-dotted) lines show the
crossover curves obtained from the peaks in the chiral (thermal)
susceptibilities. The dotted lines in the lower panels represent the
spinodals, and the fat dots indicate in each case the position of the end
point.}
\end{figure}

Finally, let us quote the numerical results for the behavior of the energy
density, the entropy density and the pressure as functions of the
temperature. The corresponding curves are shown in Fig.~5, where we plot
the scaled quantities $\varepsilon/T^4$, $\frac{3}{4}\, s/T^3$ and
$3p/T^4$ versus the relative temperature $T/T_c$, for Sets I and II and
different values of the chemical potential. The arrows indicate the value
corresponding to a free fermion system in the large $T$ limit, given by
$7\pi^2/10$ ---see Eq.~(\ref{omegafree}). Notice that for Set I there is a
range of temperatures below $T_c$ for which all three quantities are
negative. This might be taken as a further indication that for this
parameter set there is a sort of ``confinement'': somewhat below $T_c$
hadronic degrees of freedom ought to be included, and the system cannot be
simply treated as a quark gas. This represents a qualitative difference
between Sets I and II. It is interesting to see that, in spite of this
fact, the behavior at the transition region is relatively similar in both
cases.

\begin{figure}[htb]
\centerline{
   \includegraphics[height=17truecm]{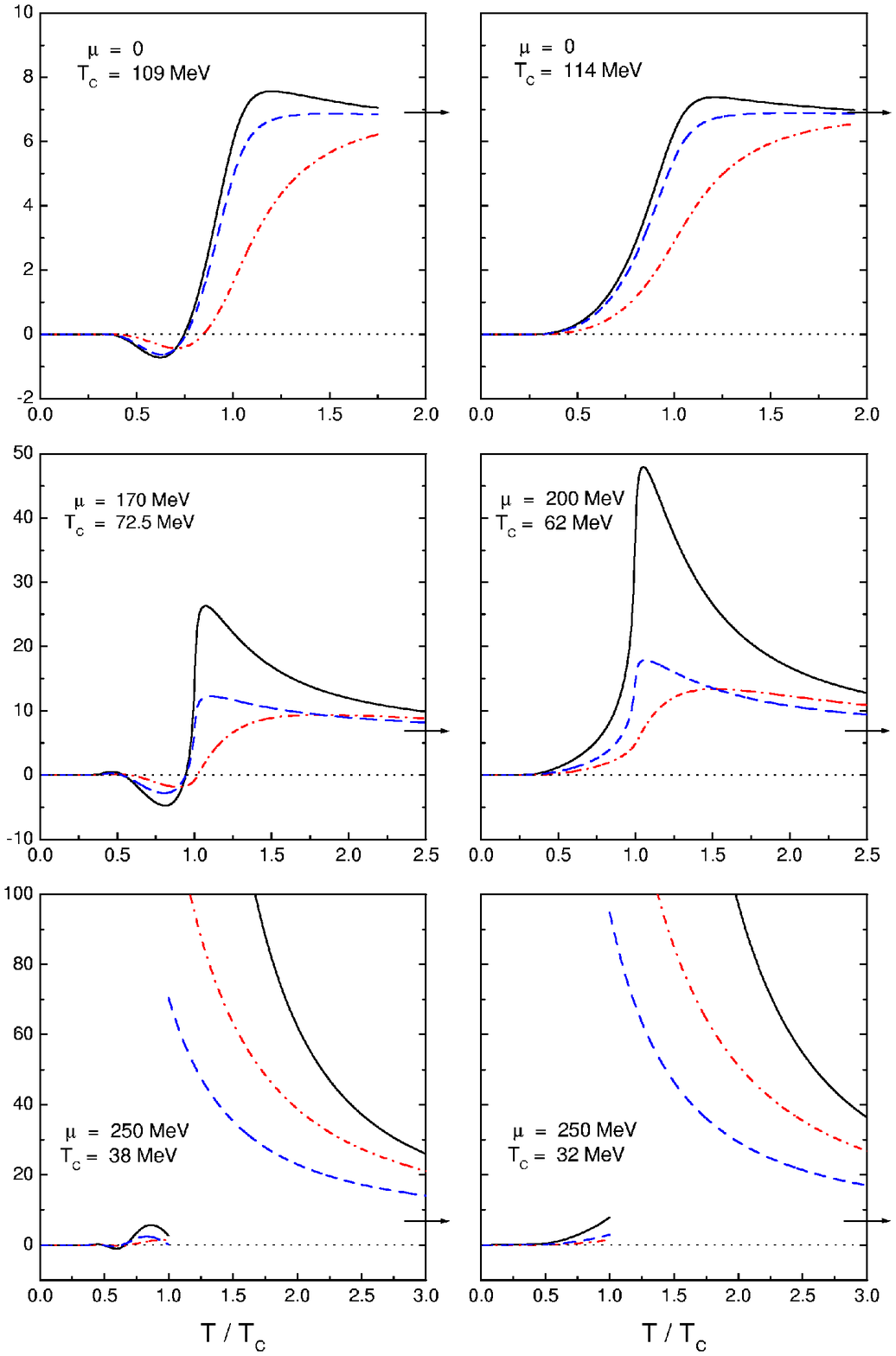}
   } \caption{Numerical results for the scaled quantities
$\varepsilon/T^4$ (solid), $\frac34 s/T^3$ (dashed) and $3p/T^4$
(dashed-dotted). Curves are given as functions of $T/T_c$, for fixed
values of $\mu$ that are indicated in each case together with the
respective critical temperatures. The arrows show the asymptotic values,
which correspond to a free fermion system. Left and right panels show the
results obtained for Sets I and II, respectively.}
\end{figure}

\section{Summary and outlook}

In this work we have presented further details of the chiral phase
transition within chiral quark models including nonlocal interactions in
the mean field approximation. In the chiral limit, our analysis allows to
obtain semi-analytical expressions for the transition curve and the
location of the critical points. For the case of finite current quark
masses, we have studied the behavior of various thermodynamical and chiral
response functions across the phase transition. In the crossover region
the thermal and chiral susceptibilities display clear peaks which allow to
define a transition temperature, in the same way as it is usually done in
lattice calculations.

The resulting phase diagrams are qualitatively similar to those obtained
within other frameworks, as e.g.\ the standard NJL model, the Bag model,
and lattice QCD. In the nonlocal schemes studied here, however, the
transition temperature at $\mu=0$ is found to be somewhat lower than the
expected value $T_c(0) \simeq 170$ MeV once the model parameters are
fitted so as to reproduce both the empirical values of $m_\pi$ and $f_\pi$
and the phenomenological value of the light quark condensates at zero
temperature and density. In fact, this result may be improved by
considering alternative ways of including nonlocality, as e.g.\ regulator
schemes inspired in one-gluon exchange processes~\cite{Sch94}, or through
the inclusion of beyond MFA contributions. Work in this direction is
currently in progress.

\section*{Acknowledgements}

We thank M.\ Malheiro for useful discussions. This work has been partially
supported by CONICET and ANPCyT, under grants PIP 02368, PICT00-03-08580
and PICT02-03-10718.

\section*{Appendix A: Expansion of the thermodynamical potential in powers
of $\qq$}

\newcounter{erasmo}
\renewcommand{\thesection}{\Alph{erasmo}}
\renewcommand{\theequation}{\Alph{erasmo}.\arabic{equation}}
\setcounter{erasmo}{1} \setcounter{equation}{0} 

The grand canonical thermodynamical potential defined by Eq.~(\ref{omega})
turns out to be divergent. Here it has been regularized by subtracting the
expression corresponding to a system of free fermions of mass $m_c$, and
adding it in a regularized form. We have thus
\begin{equation}
\omega_\mf^{(reg)} (T,\mu,m_c) = \frac{\bar \sigma^2}{2 G}\;
-\ 4 N_c \sint \
\log \left[ \frac{p^2 + \Sigma^2(p^2)}{p^2 + m_c^2} \right] +
\omega_{free}^{(reg)}(T,\mu,m_c) + \omega_0 \ ,
\label{omegreg}
\end{equation}
where
\begin{equation}
\omega_{free}^{(reg)}(T,\mu,m_c) = -\,4\,N_c\,T
\int \frac{d^3p}{(2\pi)^3}\;\left[\,\log\left(1+e^{-(E-\mu)/T}\right)
+\; \log\left(1+e^{-(E+\mu)/T}\right)\right]\;,
\label{ofreeint}
\end{equation}
with $E=\sqrt{p^2+m_c^2}\,$, while $\omega_0$ is a constant fixed by
imposing that the thermodynamical potential vanishes at zero $T$ and
$\mu$. We point out that this regularization prescription relies on the
well-behaved shape of the regulator in $\Sigma(p^2)$. Alternative schemes
can be applied in other cases, see e.g.~\cite{Bla97}.

Let us assume that in the chiral limit, $m_c=0$, the order parameter $\qq$
vanishes at a given temperature $T=T_c$. We are interested in the
description of the situation in the vicinity of this temperature, hence it
is useful to carry out a double expansion of $\omega_\mf^{(reg)}$ in
powers of $\qq$ and $m_c$. For $m_c=0$, the expansion of
$\omega_\mf^{(reg)}$ in powers of $\qq$ can be performed by taking into
account the derivatives
\begin{eqnarray}
\left. \frac{\partial^2 \omega_\mf}{\partial\qq^2}\;\right|_{\qq=0} & = &
\left. \frac{\partial^2 \omega_\mf}{\partial\sigma^2}
\left(\frac{\partial \qq}{\partial\sigma}\right)^{-2}\,\right|_{\qq=0}
\nonumber \\
\left. \frac{\partial^4 \omega_\mf}{\partial\qq^4}\;\right|_{\qq=0} & = &
\left. \frac{\partial^4 \omega_\mf}{\partial\sigma^4}
\left(\frac{\partial \qq}{\partial\sigma}\right)^{-4}
- \ 4\ \frac{\partial^2 \omega_\mf}{\partial\sigma^2}
\left(\frac{\partial \qq}{\partial\sigma}\right)^{-5}
\frac{\partial^3 \qq}{\partial\sigma^3}\;\right|_{\qq=0} \nonumber \\
\left. \frac{\partial \omega_\mf}{\partial\qq}\;\right|_{\qq=0} & = &
\left. \frac{\partial^3 \omega_\mf}{\partial\qq^3}\;\right|_{\qq=0} \ = \
\left. \frac{\partial^5 \omega_\mf}{\partial\qq^5}\;\right|_{\qq=0} \ = \ 0
\ \ .
\label{deriv}
\end{eqnarray}
Since $\qq=0$ implies $\sigma=0$, it is easy to see that the above partial
derivatives are given by
\begin{eqnarray}
\left. \frac{\partial \qq}{\partial\sigma}\;\right|_{\sigma=0}
=  -\, 4 \, N_c\ S_{11}(T,\mu) \qquad & ; & \qquad
\left. \frac{\partial^3 \qq}{\partial\sigma^3}\;\right|_{\sigma=0}
 =  24 \, N_c\ S_{32}(T,\mu) \ \ \ ;\nonumber \\
\left. \frac{\partial^2 \omega_\mf}{\partial\sigma^2}\;\right|_{\sigma=0}
= \frac{1}{G}\, - \, 8\, N_c\ S_{21}(T,\mu)  & ; & \qquad
\left. \frac{\partial^4 \omega_\mf}{\partial\sigma^4}\;\right|_{\sigma=0}
 =  48\, N_c\ S_{42}(T,\mu)\ \ \ ,
\label{integs}
\end{eqnarray}
where the functions $S_{mn}(T,\mu)$ are defined by
\begin{equation}
S_{mn}(T,\mu)\ =\ \sint \frac{r^{2m}(p^2)}{p^{2n}}\ \ .
\label{smn}
\end{equation}

On the other hand, for $m_c\neq 0$ one has
\begin{equation}
\left. \frac{\partial\omega_\mf}{\partial m_c}\;\right|_{m_c=0}
= 2\;\qq\ .
\end{equation}
Taking into account this relation, together with Eqs.~(\ref{integs}) and
(\ref{deriv}), one easily arrives at the result shown in
Eq.~(\ref{expansion}).

\section*{Appendix B: Evaluation of Matsubara sums}

\newcounter{erasmo2}
\renewcommand{\thesection}{\Alph{erasmo2}}
\renewcommand{\theequation}{\Alph{erasmo2}.\arabic{equation}}
\setcounter{erasmo2}{2} \setcounter{equation}{0} 

In this Appendix we will show how to work out the Matsubara sums $S_{m1}$
and $S_{m2}$, which are defined by Eq.~(\ref{smn}). These sums appear in
the Landau expansion of the thermodynamical potential near the critical
temperature, see Eq.~(\ref{expansion}).

In order to carry out the calculations, we take into account the analysis
in Ref.~\cite{DS02}, where Cauchy's theorem was used to convert the
Matsubara sums into an integral plus a sum over pole residues. In the case
of a function $F(p^2)$ having only simple poles and no cuts in the complex
plane one obtains~\cite{DS02}
\begin{eqnarray}
\sint \ F(p^2) & = &
\int \frac{d^4p}{(2\pi)^4}\ F(p^2)\
\nonumber \\
 & & + \ \ 2 \int \frac{d^3p}{(2\pi)^3}\ \mbox{Re}\!\!\!\!
\sum_{\scriptsize
\begin{array}{cc}
R_k > - \mu \\
I_k \geq 0
\end{array}} \!\! \gamma_k \
\mbox{Res}\,[{\cal F}(z);z_k]\,
\left[\, n_+(z_k+\mu) + n_-(z_k+\mu) \, \right]\, ,
\label{main}
\end{eqnarray}
where the function ${\cal F}$ is defined as ${\cal F}(z)\equiv
F[(-iz-i\mu)^2+|\vec p|^2]$, $z_k=R_k+i I_k$ are the residues of this
function in the complex plane $z$, and the coefficient $\gamma_k$ is
defined as $\gamma_k=1/2$ for $I_k=0$ and $\gamma_k=1$ otherwise. We have
also introduced here (complex) occupation number functions $n_\pm(z)$,
defined by
\begin{equation}
n_\pm(z) = \frac{1}{1+\exp \left[(z \mp \mu)/T\right]}\ \ .
\end{equation}

In the case of the sum $S_{21}(T,\mu)$, the corresponding function ${\cal
F}_{21}(z)$ is given by
\begin{equation}
{\cal F}_{21}(z) \ =\
\frac{r_\Lambda^{4}[-(z+\mu)^2+|\vec p|^2]}{-(z+\mu)^2+|\vec p|^2}\ \ ,
\end{equation}
which has only two simple poles located at $z^\pm=-\mu \pm|\vec p|$, with
residues $\mp (2|\vec p|)^{-1}$. Therefore, Eq.~(\ref{main}) can in
principle be applied. There is, however, an subtle point to be taken into
account. The derivation of Eq.~(\ref{main}) assumes that $|{\cal F}(z)|
\to 0$ when ${\rm Re\ } z\to \infty$, which is in fact true for all the
situations considered in Ref.~\cite{DS02}. However, depending on the
specific form of the regulator, this condition might be not satisfied by
the function ${\cal F}_{21}(z)$. If this is the case, Eq.~(\ref{main}) can
still be applied provided that $T, \mu \ll \Lambda$ (see below).

Assuming that Eq.~(\ref{main}) holds, one gets
\begin{equation}
S_{21}(T,\mu) \ = \ \sint  \frac{r_\Lambda^{4}(p^2)}{p^2}\ =
\ \frac{1}{8\pi^2}\int_0^\infty \ dp\ p\ r_\Lambda^{4}(p^2)
- \ \frac{1}{4\pi^2}\int_0^\infty \ dp\ p\ [\, n_+(p) + n_-(p)\,]\ .
\label{sm1}
\end{equation}
The last integral in this expression can be worked out analytically. One
has
\begin{equation}
\int_0^\infty \ dp\ p\ [\, n_+(p) + n_-(p)\,]\ = \ -\; T^2 \left[
{\rm Li}_2(-e^{\mu/T})\; +\; {\rm Li}_2(-e^{-\mu/T}) \right] \
= \ \frac{\pi^2\; T^2}{6}\ + \frac{\mu^2}{2} \ \ ,
\label{integ}
\end{equation}
which together with Eq.~(\ref{sm1}) leads to the result quoted in
Eq.~(\ref{second}). As we have mentioned, this relations are only valid
for sufficiently low values of $T, \mu$ compared with the cutoff scale
$\Lambda$. In the case of the Gaussian regulator we have checked that for
$T \leq \Lambda/6$ and $\mu \leq \Lambda/4$ Eq.~(\ref{second}) is verified
with an accuracy of less that 1\%. In the case of the Lorentzian regulator
the region of validity is somewhat smaller, but at the same time the
relevant cutoff parameter is larger than in the Gaussian case~\cite{DS02}.
As a conclusion, we find that in all cases considered here
Eq.~(\ref{second}) can be taken to be valid with very good approximation
for the values of $T$ and $\mu$ of physical interest.

A similar analysis can be carried out for the sum $S_{42}(T,\mu)$. Here
the situation is somewhat more complicated, since one has to deal with
double poles and Eq.~(\ref{main}) is no longer valid. One has instead
\begin{eqnarray}
\hspace{-.5cm} S_{42}(T,\mu) & = & \sint  \frac{r^8(p^2)}{p^4}
\nonumber \\
& = & \int \frac{d^4p}{(2\pi)^4}\ F(p^2)\
+ \int \frac{d^3p}{(2\pi)^3}\ \left[
\mbox{Res}\,\left(\frac{{\cal F}(z)}{1+e^{z/T}}\right)_{z^+} \!\!
- \mbox{Res}\,\left(\frac{{\cal F}(z)}{1+e^{-z/T}}\right)_{z^-}
\right]\;,
\end{eqnarray}
where $F(p^2)=r_\Lambda^{8}(p^2)/p^4$, ${\cal F}(z)=F(-(z+\mu^2)+|\vec
p|^2)$. Evaluating the residues, this leads to
\begin{eqnarray}
\hspace{-.5cm}S_{42}(T,\mu) & = & \frac{1}{8\pi^2}\int_0^\infty \ dp\
\bigg\{\frac{1}{2\pi}\left(\int_{-\infty}^\infty dq \
\frac{r_\Lambda^{8}(p^2+q^2)}{(p^2+q^2)^2}\right) -
\frac{1}{p} \left[n_+(p)+n_-(p)\,\right] \nonumber \\
& & - \ 4\, m\ r_\Lambda'(0)\ p\, [\, n_+(p) + n_-(p)\,]\
- \ \frac{1}{T}\, [\, n_+(p) + n_-(p) - n_+^2(p) - n_-^2(p)\,]
\bigg\}\; ,
\label{sm2}
\end{eqnarray}
where $r_\Lambda'(0)=dr_\Lambda(p^2)/dp^2|_{p^2=0}\;$. The first two terms
lead to divergent integrals that can be worked out with the help of a
definite regulator, e.g.\ the Gaussian one, writing
\begin{equation}
r_\Lambda^{8}(x^2) = r_\Lambda^{8}(x^2) - e^{-4 x^2/\Lambda^2} + e^{-4 x^2/\Lambda^2}\;.
\end{equation}
For the Gaussian regulator, the sum of the divergent integrals in
(\ref{sm2}), properly regularized, is given by
\begin{eqnarray}
\int_0^\infty \ dp\
\left\{\frac{1}{2\pi}\left(\int_{-\infty}^\infty dq \
\frac{e^{-4(p^2+q^2)/\Lambda^2}}{(p^2+q^2)^2}\right) -
\frac{1}{p} \left[n_+(p)+n_-(p)\,\right] \right\} & = & \nonumber \\
& & \hspace{-10cm} = \ \ \frac12 + \frac\gamma2 - \log(\pi\sqrt{m})
- \log\left(T/\Lambda\right) - f(\mu/T)\; ,
\label{reg}
\end{eqnarray}
where
\begin{equation}
f(x)\ =\ 2\; \sinh^2(x/2) \int_0^\infty \frac{dy}{y}\ \frac{\tanh
(y/2)}{\cosh x + \cosh y}\ \ .
\label{fx}
\end{equation}
Finally, the integral of the last term in Eq.~(\ref{sm2}) can be
explicitly performed, giving simply
\begin{equation}
\int_0^\infty\ dp\ [\;n_+(p) + n_-(p) - n_+^2(p) - n_-^2(p)\;]\ =\ T \ .
\label{ult}
\end{equation}
The previous result in Eq.~(\ref{integ}), together with Eqs.~(\ref{sm2})
to (\ref{ult}), lead to the expression quoted in Eq.~(\ref{s42}). Again
this relation is strictly valid for $T, \mu \ll \Lambda$. For the Gaussian
regulator we have checked that it still holds up to 1\% in the region $T
\leq \Lambda/9$, $\mu \leq \Lambda/5$. As before, this is also the case
for the Lorentzian regulator for the values of $T$ and $\mu$ relevant for
our analysis.

It is interesting to point out the similarity between our equations
determining the second order transition line and the tricritical point,
and those obtained in Ref.~\cite{zsolt}. In that article, the authors
address the study of the $T-\mu$ phase diagram using a different approach,
in which they propose a large flavor number expansion and a resummed
renormalization scheme. It can be seen that the results in
Ref.~\cite{zsolt} ---Eqs.~(13) and (14)--- have the same form as our
expressions if one takes the limit $\lambda=0$, which in the context of
Ref.~\cite{zsolt} means to neglect the contributions from the meson sector
(this should be analogous to the MFA considered here). Moreover, the
validity of these results is also limited to relatively low values of $T$
and $\mu$ in view of the presence of a Landau pole arising from the
renormalization procedure.


\begin{thebibliography}{999}

\bibitem{All03}
C.R.~Allton {\em et al.}, Phys.~Rev.~D {\bf 68}, 014507 (2003).

\bibitem{Fod04}
 Z.~Fodor and S.D.~Katz, JHEP {\bf 0203}, 014
(2002); JHEP {\bf 0404}, 050 (2004).

\bibitem{Kar03} F.~Karsch and E.~Laermann,
arXiv:hep-lat/0305025.

\bibitem{GDS00}
I.\ General, D.\ G\'omez Dumm, N.N.\ Scoccola,
Phys.\ Lett.\ B {\bf 506}, 267 (2001).

\bibitem{DS02}
D.\ G\'omez Dumm, N.N.\ Scoccola,
Phys.\ Rev.\ D {\bf 65}, 074021 (2002).

\bibitem{Rip97}
G.\ Ripka,
{\it Quarks bound by chiral fields}
(Oxford University Press, Oxford, 1997).

\bibitem{reports}
U.\ Vogl and W.\ Weise,
Prog.\ Part.\ Nucl.\ Phys.\ {\bf 27}, 195 (1991);
S.\ Klevansky,
Rev.\ Mod.\ Phys.\ {\bf 64}, 649 (1992);
T.\ Hatsuda and T.\ Kunihiro,
Phys.\ Rep.\ {\bf 247}, 221 (1994).

\bibitem{SS98}
T.\ Schafer and E.V.\ Schuryak,
Rev.\ Mod.\ Phys.\ {\bf 70}, 323 (1998).

\bibitem{RW94}
C.D.\ Roberts and A.G.\ Williams,
Prog.\ Part.\ Nucl.\ Phys.\ {\bf 33}, 477 (1994);
C.D.\ Roberts and S.M.\ Schmidt,
Prog.\ Part.\ Nucl.\ Phys.\ {\bf 45}, S1 (2000).

\bibitem{AS99}
E.\ Ruiz Arriola and L.L.\ Salcedo,
Phys.\ Lett.\ B {\bf 450}, 225 (1999).

\bibitem{Rip00}
G.\ Ripka,
Nucl.\ Phys.\ A {\bf 683}, 463 (2001);
R.S.\ Plant and M.C.\ Birse,
Nucl.\ Phys.\ A {\bf 703}, 717 (2002).

\bibitem{Sti87} M.~Stingl, Phys.\ Rev.\ {\bf D 34}, 3863 (1986)
[Erratum-ibid.\ {\bf D 36}, 651 (1987)];
H.J.~Munczek, Phys.\ Lett.\ {\bf B 175}, 215 (1986);
C.J.~Burden, C.D.~Roberts and A.G.~Williams, Phys.\ Lett.\ {\bf B 285},
347 (1992);
G.~Krein, C.D.~Roberts and A.G.~Williams, Int.\ J.\ Mod.\ Phys.\
{\bf A 7}, 5607 (1992);
D.~Blaschke, G.~Burau, Y.L.~Kalinovsky, P.~Maris and P.C.~Tandy, Int.\
J.\ Mod.\ Phys.\ A {\bf 16}, 2267 (2001).

\bibitem{BB95}
R.D. Bowler and M.C. Birse,
Nucl. Phys. A {\bf 582}, 655 (1995);
R.S.Plant and M.C. Birse,
Nucl. Phys. A {\bf 628}, 607 (1998).

\bibitem{SDS04}
A. Scarpettini, D. G\'omez Dumm and N.N. Scoccola,
Phys. Rev. {\bf D69}, 114018 (2004).

\bibitem{BGR02}
W.~Broniowski, B.~Golli and G.~Ripka,
Nucl. Phys. {\bf A703}, 667 (2002);
A.H. Rezaeian, N.R. Walet and M.C. Birse,
Phys.\ Rev.\ C {\bf 70}, 065203 (2004).

\bibitem{DGS04}
R.S. Duhau, A.G. Grunfeld and N.N. Scoccola,
Phys.\ Rev.\ D {\bf 70}, 074026 (2004).

\bibitem{SKP99}
T.M.~Schwarz, S.P.~Klevansky and G.~Papp, Phys.\ Rev.\ C {\bf 60},
055205 (1999).

\bibitem{khuang}
See e.g.\ K.\ Huang, {\it Statistical Mechanics}, (J. Wiley \& Sons, New
York, 1987), Sec.\ 17.4.

\bibitem{expo}
D.~Blaschke, A.~Holl, C.D.~Roberts and S.M.~Schmidt,
Phys.\ Rev.\ C {\bf 58}, 1758 (1998);
A.~Holl, P.~Maris and C.D.~Roberts,
Phys.\ Rev.\ C {\bf 59} 1751 (1999).

\bibitem{zsolt}
A.~Jakov\'ac, A.~Patk\'os, Zs.~Sz\'ep and P.~Sz\'epfalusy, Phys.\ Lett.\ B
{\bf 582}, 179 (2004).

\bibitem{BR98}
J. Berges and K. Rajagopal, Nucl.\ Phys.\ B {\bf 538}, 215 (1999).

\bibitem{Bub04}
M.~Buballa, Phys.\ Rept.\  {\bf 407}, 205 (2005).

\bibitem{Sch94}
S.M.~Schmidt, D.~Blaschke and Y.L.~Kalinovsky, Phys.\ Rev.\ C {\bf 50},
435 (1994).

\bibitem{Bla97}
D.~Blaschke, C.D.~Roberts and S.M.~Schmidt, Phys.\ Lett.\ B {\bf 425},
232 (1998).

\end{thebibliography}
\end{document}